\documentclass{JINST}
\usepackage{graphicx, subfigure}

\title{Development of a modular test system for the silicon sensor R\&D of the ATLAS Upgrade}

\author{H. Liu,$^{a,b,}$\thanks{Corresponding
		author.}~~M. Benoit,$^c$ H. Chen,$^b$ K. Chen,$^b$ F. A. Di Bello,$^c$ G. Iacobucci,$^c$ F. Lanni,$^b$ I. Peric,$^d$ B. Ristic,$^{c,e}$ M. Vicente Barreto Pinto,$^c$ W. Wu,$^b$ L. Xu$^b$~and G. Jin$^a$\\
	\llap{$^a$}
	State Key Laboratory of Particle Detection and Electronics,\\
	University of Science and Technology of China,\\	
	Hefei, Anhui 230026, China\\
	\llap{$^b$}Brookhaven National Laboratory,\\
	Physics Department,\\
	Upton, NY 11973, USA\\
	\llap{$^c$}University of  Geneva,\\
	Department of Nuclear and Particle Physics, \\
	Geneva, Switzerland\\
	\llap{$^d$}Karlsruhe Institute of Technology (KIT),\\
	Kaiserstra$\beta$e 12,\\
	76131 Karlsruhe, Germany\\
	\llap{$^e$}European Organization for Nuclear Research (CERN),\\
	385 Route de Meyrin,\\
	1217 Meyrin, Switzerland\\
	E-mail: \email{hliu2@bnl.gov}}

\abstract{High Voltage CMOS sensors are a promising technology for tracking detectors in collider experiments. Extensive R\&D studies are being carried out by the ATLAS Collaboration for a possible use of HV-CMOS in the High Luminosity LHC upgrade of the Inner Tracker detector. CaRIBOu (Control and Readout Itk BOard) is a modular test system developed to test Silicon based detectors. It currently includes five custom designed boards, a Xilinx ZC706 development board, FELIX (Front-End LInk eXchange) PCIe card and a host computer. A software program has been developed in Python to control the CaRIBOu hardware. CaRIBOu has been used in the testbeam of the HV-CMOS sensor \mbox{AMS180v4} at CERN. Preliminary results have shown that the test system is very versatile. Further development is ongoing to adapt to different sensors, and to make it available to various lab test stands.}

\begin{document}

	
\section{Introduction}
\label{sec:intro}
The ATLAS \cite{atlasexp} experiment is planning to build and install a new all-silicon Inner Tracker (ITk) for the High-Luminosity LHC (HL-LHC) upgrade. Extensive R\&D on sensors based on High Voltage CMOS (HV-CMOS) processes is on-going, given the potential multiple advantages of this technology compared to the traditional planar pixel detectors \cite{hvcmos}. Several prototypes with different pixel types have been designed and manufactured in the 180 nm and 350 nm HV-CMOS processes provided by Austria Mikro Systeme (AMS) \cite{amshvcmos}. Radiation hardness and performance of these detectors need to be evaluated before they are qualified for the HL-LHC upgrade. A versatile test system, which can be easily adapted to different types of sensors, is necessary. CaRIBOu (Control and Readout Itk BOard) is a modular test system developed for the HV-CMOS sensor R\&D studies.\par

In this paper, overview of the CaRIBOu system is described in section~\ref{sec:overview}. Section~\ref{sec:hw} provides the details of the hardware design, which is followed by the Xilinx ZC706 development board \cite{zc706} based FPGA firmware design and host software design in sections~\ref{sec:fw} and \ref{sec:sw}. After that, the threshold tuning method utilized and result of the pixel read out chip and one pixel sensor are presented in section~\ref{sec:fei4_tune} and section~\ref{sec:ccpd_tune}. Finally, the utilization of the CaRIBOu system in testbeam is summarized in section~\ref{sec:testbeam}.

\section{System Overview}
\label{sec:overview}
\begin{figure}[htbp]
\centering 
\includegraphics[width=1\textwidth]{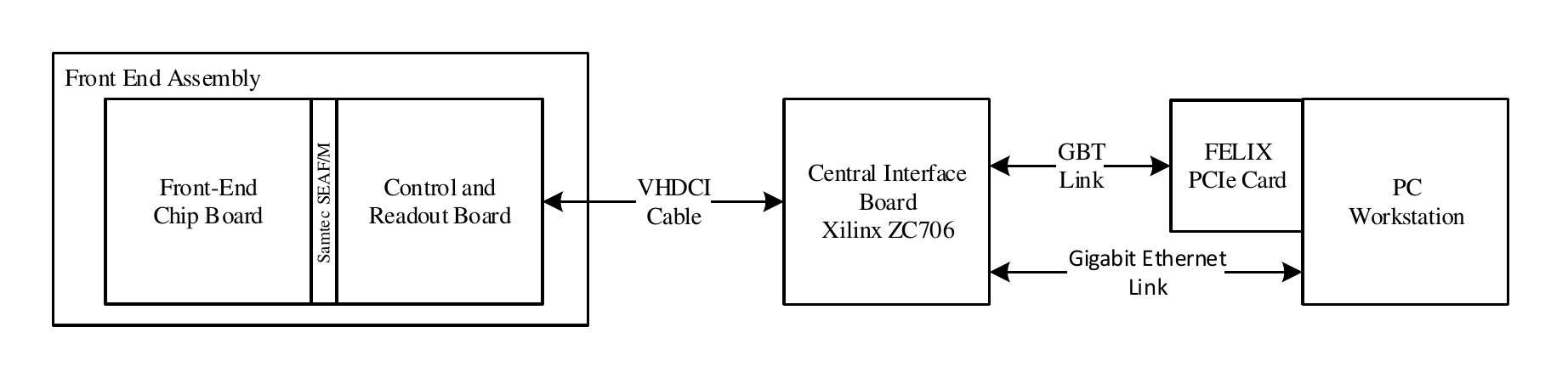}
\caption{\label{fig:sys_bd}Block Diagram of the CaRIBOu system.}
\end{figure}

Figure~\ref{fig:sys_bd} shows the block diagram of the CaRIBOu system for silicon sensor evaluation test. The components of this system can be classified into two categories: front-end and back-end. The front-end components are connected to the back-end by the VHDCI (Very High Density Cable Interconnect) cable \cite{molexvhdci} with the help of VHDCI adapter cards.\par

Front-end assembly includes one control and readout (CaR) board and several front-end chip boards.
The CaR board is designed to provide all the necessary resources for the readout chip and silicon sensor under test. The CaR board includes adjustable power supplies with monitoring capability, bias voltages, configuration LVCMOS signals, calibration injection pulse, and analog input channels of ADC. Samtec SEARAY \cite{searaycon} connectors, with features of small footprint and high density, are selected to connect the CaR board and front-end chip boards.
Front-end chip boards are sensor and readout chip specific boards. The front-end readout chip and the silicon sensor under test are mounted on and wire bonded to the front-end chip boards. \par

Back-end comprises a central interface board, a FELIX \cite{felix} PCIe card and a PC workstation.
The central interface board is the core of the CaRIBOu system, which is based on the Xilinx ZC706 FPGA development board. This board is used to control the power rails, bias voltages, pulse generators of the CaR board, to work as the ADC receiver and controller, and to configure the front-end readout chip and silicon sensor according to the commands issued by the PC workstation. In addition, the data from the front-end readout chip are decoded, packaged and sent to the PC workstation through the Gigabit Ethernet link or the optical GBT-Link for off-line analysis. The former one is more convenient for the lab test and development, while the latter one is more close to the final readout architecture of the ATLAS experiment.

\section{Hardware Design}
\label{sec:hw}
Five custom PCB boards have been designed for this modular test system, including one CaR board, two front-end chip boards and two VHDCI adapter cards.

\subsection{Control and Readout (CaR) Board}
\begin{figure}[htbp]
\centering 
\includegraphics[width=.4\textwidth,angle=270]{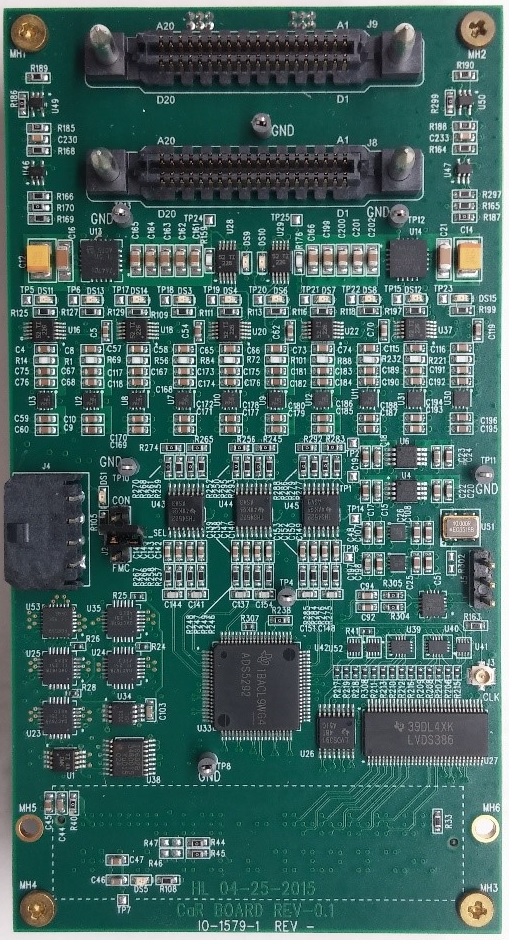}
\caption{\label{fig:car_board} Top side of the Control and Readout (CaR) board.}
\end{figure}

The CaR board is designed as a FMC mezzanine card \cite{vita57} with a FMC LPC connector, shown in Figure~\ref{fig:car_board}. It has two Samtec SEARAY 80 pin male connectors for the connection of the front-end chip boards. The CaR board can provide a total of eight adjustable power supplies with a maximum current capability of 500 mA, and two fixed 1.8 V power supplies with a 1A maximum output current to the front-end chip boards. All these power rails can be monitored through the I2C current monitoring chip INA226 \cite{ina226} from Texas Instruments. In addition, sixteen bias voltages with a maximum driver strength of 20 mA, a 40 MHz 12 bit ADC with 8 analog input channels, and two calibration pulse generators can be accessed by the front-end chip boards. All LVDS signals from the central interface board, except those assigned for the front-end readout chip \mbox{FE-I4B} \cite{fei4ref1} \cite{fei4ref2}, are converted to LVCMOS signals, and fed to the front-end chip boards for detector configuration. The CaR board is a general purpose control and readout board, it is planned to be used for the test of other silicon sensors.

\subsection{Front-End Chip Boards}
\begin{figure}[htbp]
\centering 
\includegraphics[width=.7\textwidth]{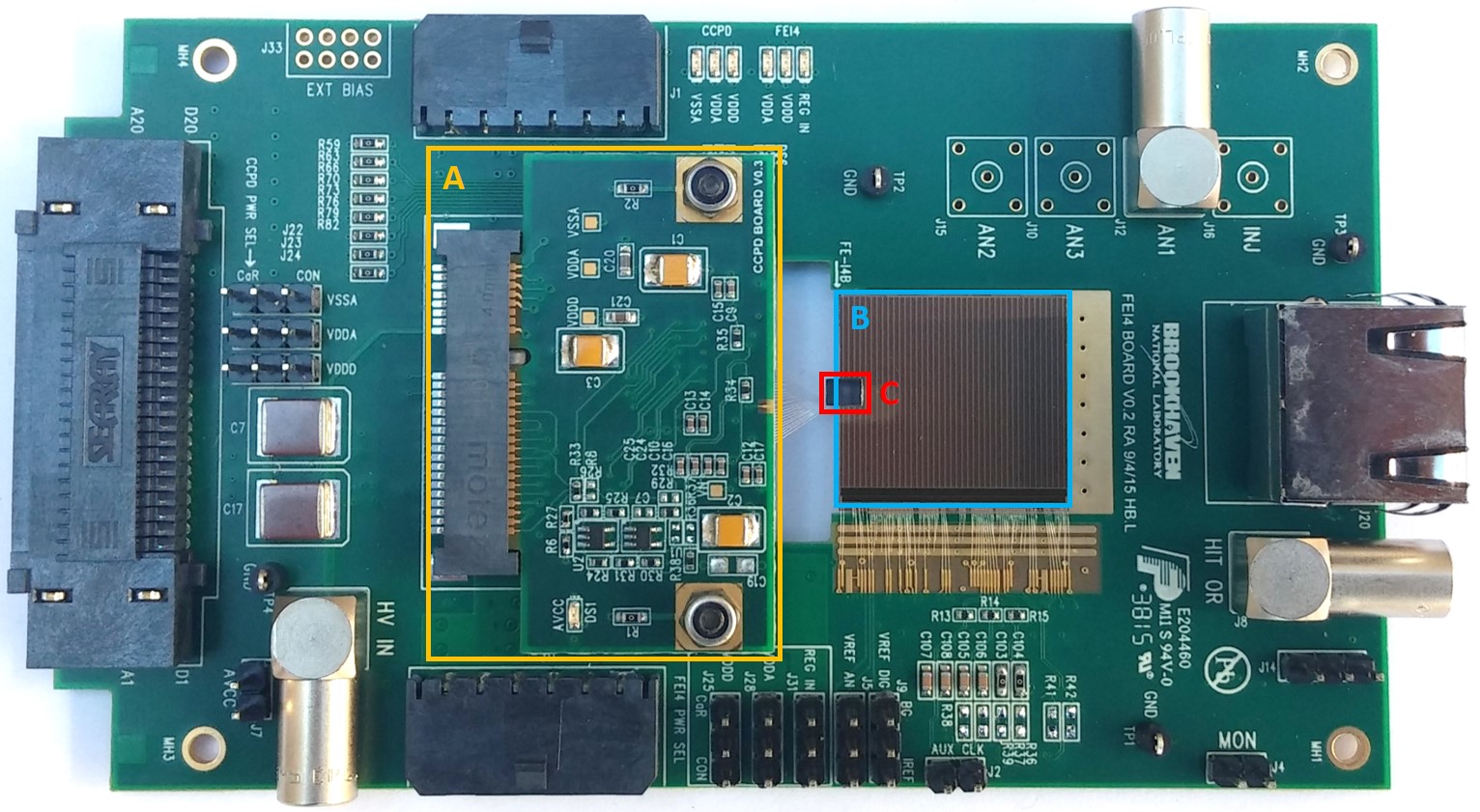}
\caption{\label{fig:fei4_board}FEI4 board with CCPD board (A) plugged in. The large chip (B) in the center is the front-end pixel readout chip \mbox{FE-I4B}. The small chip (C) glued on the \mbox{FE-I4B} is the HV-CMOS pixel sensor \mbox{AMS180v4}.}
\end{figure}

The performance of the sensor manufactured by the AMS HV-CMOS 180 nm process, which is named \mbox{AMS180v4}, is currently under investigation. For the test of AMS180v4 sensors, the \mbox{FE-I4B} ASIC is used as the front-end readout chip. 
Front-end chip boards consist a carrier board and sensor mezzanine. The carrier board is the FEI4 board with Samtec right-angle SEARAY 80-pin connector, and the sensor mezzanine is designed as a mezzanine board with mini-PCIe interface. The front-end chip boards are shown in Figure~\ref{fig:fei4_board}.\par
The FEI4 board is used to mount the assembly of the readout chip \mbox{FE-I4B} and the pixel sensor \mbox{AMS180v4}. Wire bonding pads for the bare-die \mbox{FE-I4B} chip are located on the FEI4 board as well.\par

A mini-PCIe socket is placed on the FEI4 board to connect the CCPD board. The CCPD board is designed as a small mini-PCIe card, with bonding pads for the \mbox{AMS180v4} sensor located on the bottom of this board.\par

All the configuration signals for the \mbox{AMS180v4} and \mbox{FE-I4B} are connected to the CaR board through the 80-pin SEARAY connector. 

\subsection{VHDCI Adapter Boards}
To avoid the ground loop between the central interface board and the front-end assembly, all the signals between front-end and back-end are implemented as LVDS differential signals, connected with a VHDCI cable.\par
Since the CaR board is a FMC mezzanine card, two VHDCI adapter boards have been designed. One is the FMC to VHDCI adapter board, the other is the VHDCI to FMC adapter board. A differential I2C bus is implemented on the adapter cards by placing an I2C differential buffer PCA9614 from NXP semiconductors on both boards.

\section{FPGA Firmware Design}
\label{sec:fw}
\begin{figure}[htbp]
\centering 
\includegraphics[width=0.8\textwidth]{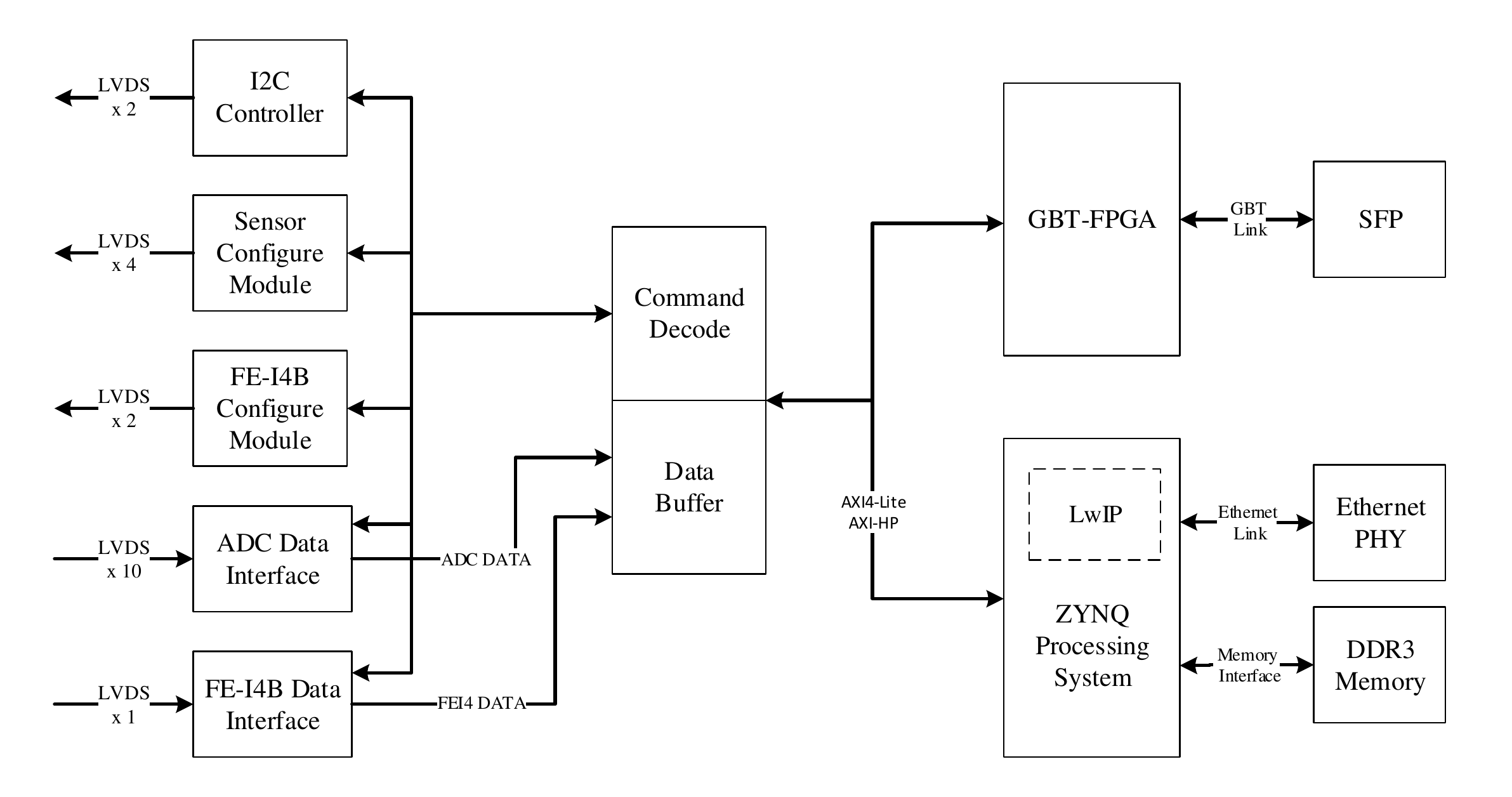}
\caption{\label{fig:firmware_bd}Block diagram of the FPGA firmware of the central interface board ZC706 for the test of the sensor \mbox{AMS180v4} with FE-I4B as its readout chip.}
\end{figure}

The block diagram of the FPGA firmware for the test of the sensor \mbox{AMS180v4} with FE-I4B as its readout chip is shown in Figure~\ref{fig:firmware_bd}. \par

There are two links that can be used to transfer the data and commands between the central interface board and the PC workstation, the Ethernet link developed with ZYNQ \cite{zynq} processing system with LwIP protocol \cite{lwip} and the GBT-link implemented with the low latency version GBT-FPGA IP core \cite{gbtfpga} \cite{low_lantency_gbtfpga}.\par

The commands from the PC workstation are transferred to the command decode module through the AXI4 bus or the custom designed IP bus, when Ethernet link or GBT link are used respectively. This command decode module is acting as the controller of all the internal logic modules.\par 

An I2C controller module is implemented to control and monitor the power rails, to adjust bias voltages, and to configure the ADC with SPI interface through an I2C to SPI bridge chip NXP SC18IS602B on the CaR board. Sensor and FE-I4B configuration modules are used to configure sensors under test and the \mbox{FE-I4B} chip. Besides these control modules, two interface modules have been developed to decode and buffer the data from the ADC and \mbox{FE-I4B} devices.  Data from the pixel readout chip \mbox{FE-I4B} is a 160 Mbps 8b/10b encoded serial signal \cite{fei4manual}, it is de-serialized, aligned, 10b/8b decoded and extracted by the FE-I4B data interface module. Then the buffered data from the FE-I4B and ADC can be sent to the host computer through the Ethernet or GBT optical link.\par

 When the Ethernet link is used, the buffered data are written into the DDR3 memory through the AXI-HP interface for Ethernet transmission. If the GBT link is selected, all of these buffered data will be packaged into the 120 bit GBT-FPGA frame. The data frame then will be sent to the FELIX PCIe card through the GBT link, and stored in the hard drive of the workstation via the PCIe interface.\par
 
 \section{Software Design}
 \label{sec:sw}
 The host software is responsible for issuing all the control commands to the central interface board, and store all the pixel data from the front-end for off-line analysis. The software is developed in Python and its GUI is based on the PyQt library \cite{pyqt}. \par 
 
The software can talk to the central interface board directly through the Gigabit Ethernet link by TCP/IP socket, or through the PCIe and GBT-link by configuring the PCIe registers of the FELIX PCIe card.\par
 Users can control and monitor all the power rails, configure the readout chip \mbox{FE-I4B} and sensor under test, tune different sensor parameters and read back the front-end data etc. by the software.
 In addition, several calibration algorithms have been implemented, including the threshold tuning for the FE-I4B and AMS180v4 , and the TOT (Time Over Threshold) tuning for FE-I4B.\par  

 
\section{Application} 
Currently, the CaRIBOu system has been used for the testbeam of the HV CMOS sensor AMS180v4 and H35DEMO. The test of H35DEMO is ongoing and will not be presented here.

\subsection{FE-I4B Tuning}
\label{sec:fei4_tune}
The FE-I4B integrated circuit is used as the readout chip of the sensor AMS180v4 during the test, and the signals from the sensor are capacitive coupled to the pixels of FE-I4B. Each FE-I4 pixel has an independent, free running amplification stage with adjustable shaping, followed by a discriminator with adjustable threshold. The threshold applied to the comparator and the feedback of the amplification stage of each pixel can be adjusted by configuring the pixel FDAC (Feedback DAC) and TDAC (Threshold DAC) register respectively. \cite{fei4manual} In order to eliminate the threshold and feedback current variance between pixels, threshold tuning and TOT tuning of the FE-I4B have been performed before using it to read out the signals from the sensor.\par 

The tuning algorithms for the FE-I4B are implemented in the software, by issuing calibration commands and analyzing the data read back from the FE-I4B. The threshold tuning is performed first, this step can adjust the effective threshold of all the pixel close to the target threshold. After threshold tuning, the TOT tuning procedure can adjust the feedback current of every pixel to make the output TOT value of a specific input close to the preset TOT value target. The TOT tuning will affect the effective threshold, so the threshold tuning has been performed for a second time.\par

\begin{figure} 
	\centering 
	\subfigure[~]{\includegraphics[width={0.48\textwidth}]{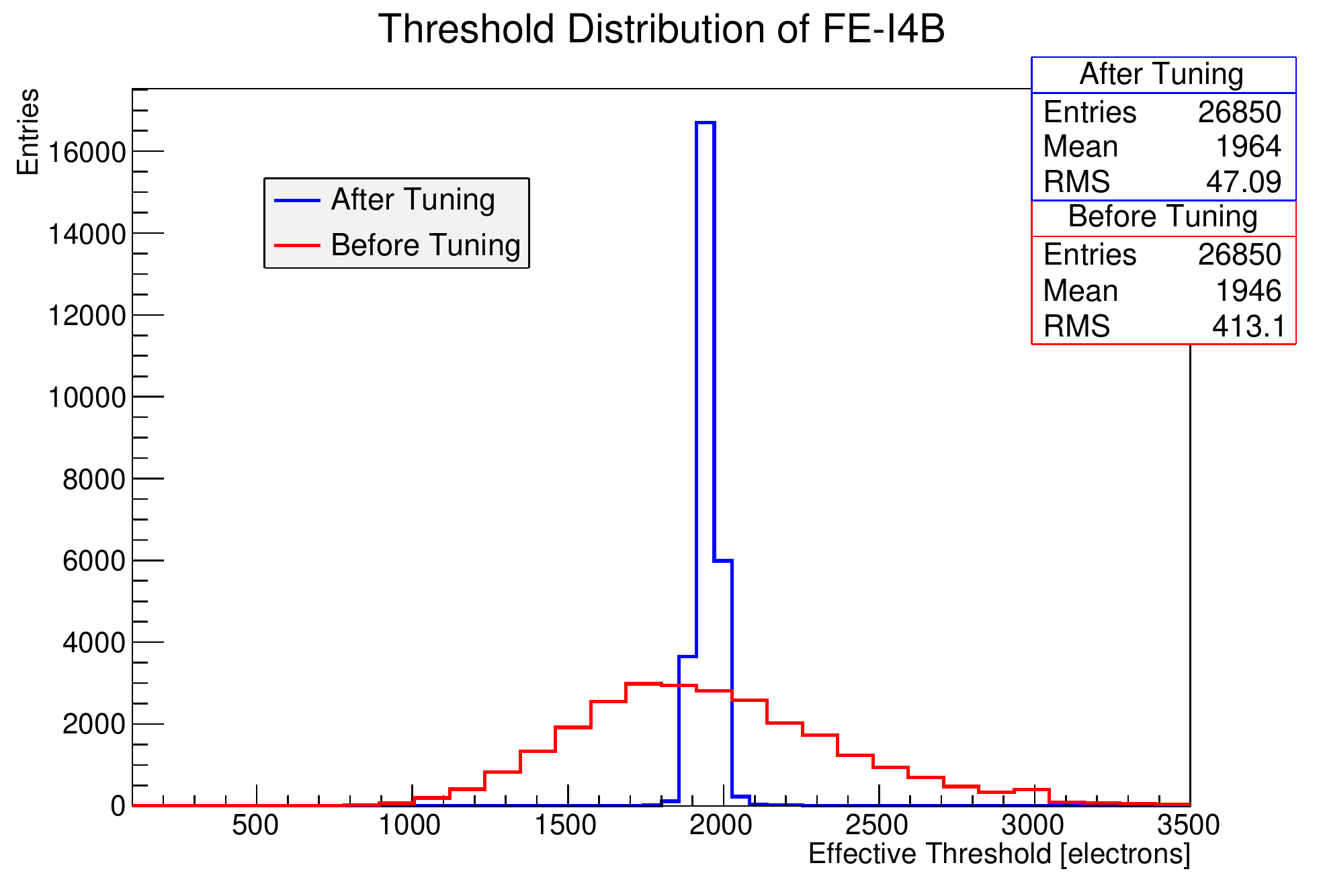}\label{fig:fei4_th_after} }
	\subfigure[~]{\includegraphics[width={0.48\textwidth}]{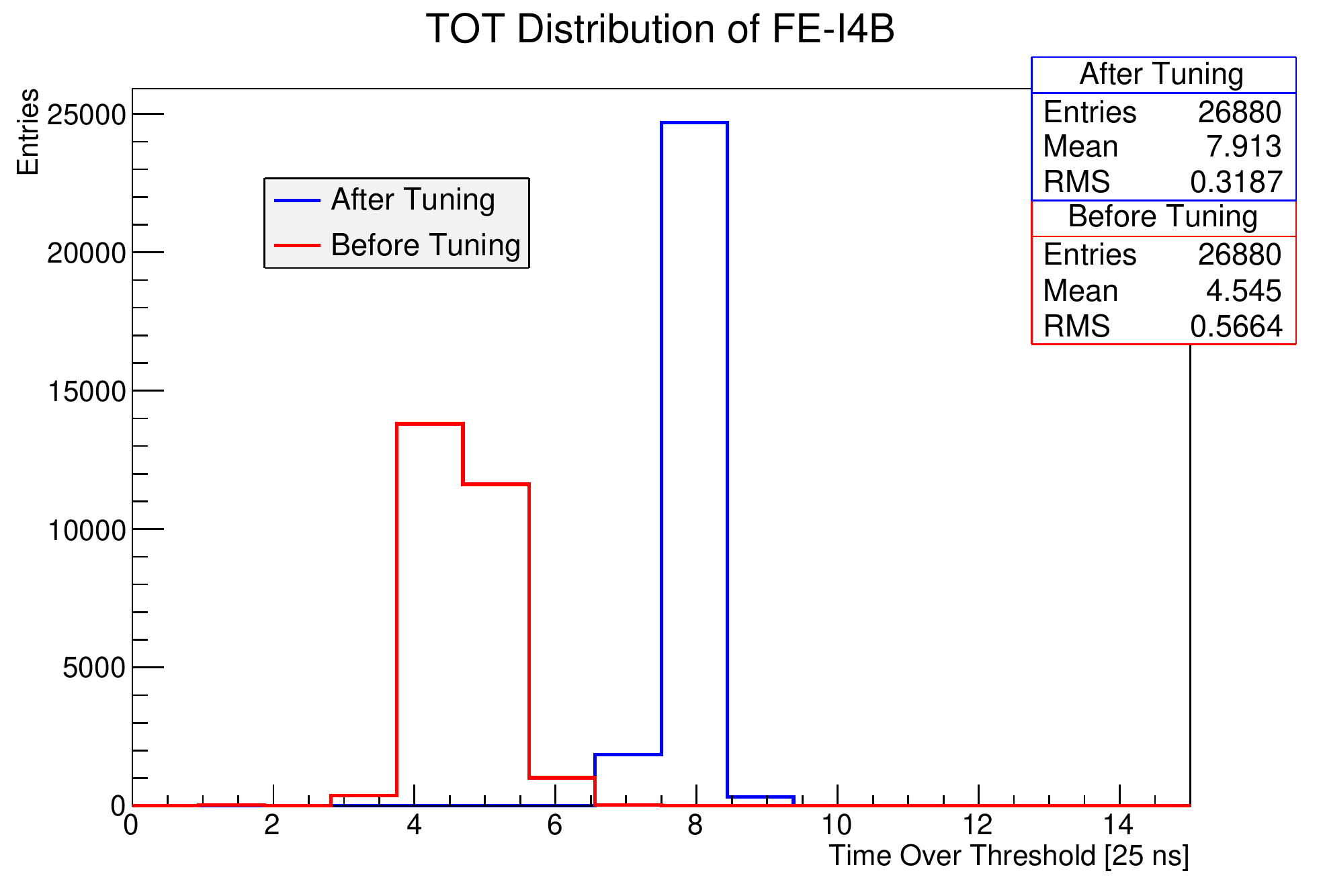} \label{fig:fei4_tot_after} } 
	\caption{Tuning results of the FE-I4B integrated circuit. (a) Effective threshold distribution before and after tuning. The target threshold is set to 2000 electrons, and the mean value of the distribution can be adjusted by changing the global DAC output voltage by configuring global register. (b) TOT distribution before and after tuning. The tuning target is set to TOT = 8 BC (Bunch Crossing) for the calibration pulses with amplitude of 20000 electrons.} 
	\label{fig:fei4_threshold_tot_dist} 
\end{figure}

The tuning results of one FE-I4B sample are showed in Figure~\ref{fig:fei4_threshold_tot_dist}, the effective threshold deviation after tuning is around 47 electrons, this is consistent to the specification of the FE-I4B \cite{fei4manual}. The CaRIBOu system can complete the threshold tuning and TOT tuning procedures for one FE-I4B within 3 minutes. \par

\subsection{Sensor Tuning}
\label{sec:ccpd_tune}
\begin{figure} 
	\centering 
	\includegraphics[width={0.6\textwidth}]{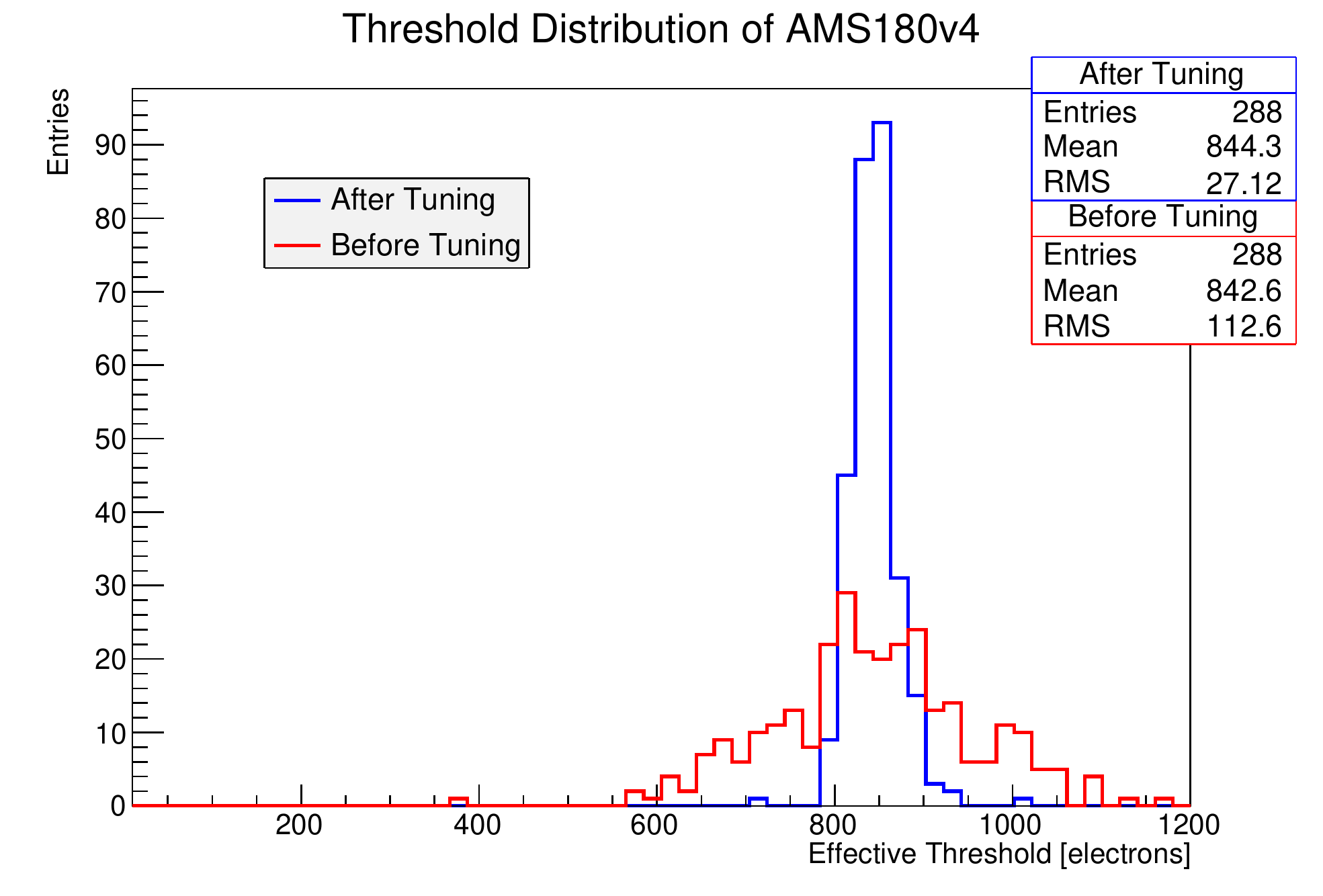}\label{fig:h18v4_th_before} 
	\caption{Effective threshold distribution of the sensor AMS180v4 before and after pixel DAC tuning. The mean value of the effective can be adjusted by change the external threshold bias voltage. } 
	\label{fig:h18v4_threshold_dist} 
\end{figure}

To ensure the good quality and uniformity of the data recorded, the detection threshold of each pixel of the sensor under test must be uniform. This also allows for reaching lower thresholds without outlier pixels contributing to the noise occupancy of the detector. For that the pixel threshold of the sensor need to be equalized.\par
 
Threshold tuning of the sensor \mbox{AMS180v4} has been performed before the testbeam. The \mbox{AMS180v4} sensor has four kinds of pixel, and only the baseline one, STime pixel, was measured. The per-pixel analogue electronic of \mbox{AMS180v4} comprises charge sensitive amplifiers, source followers and a discriminator stage which can be controlled by two voltage levels (global threshold from external and local threshold generated by the local DAC) \cite{2015testbeamresults}. \par 

The tuning procedure controls the 4-bit local DAC (named TDAC - Threshold DAC) connected to the discriminator of each pixel of the \mbox{AMS180v4} sensor to compensate the possible local mismatches in the electronics between different pixels. The TDAC can be seen as a correction to the pixel local threshold resulting in an effective threshold (when summed with the global threshold). The corrected value of the voltage for each TDAC must be selected in order to have a uniform effective-threshold value across the pixel matrix.\par

The result of the threshold tuning can be seen in Figure~\ref{fig:h18v4_threshold_dist}. The black distribution shows the effective threshold achieved after the tuning while the blue and red distributions show, respectively, the effective threshold for TDAC 1 and 14. It can be observed that the tuning procedure results in a more narrow distribution of the effective threshold with a dispersion of 27 electrons with respect to 112 electrons before tuning.\par

\subsection{Testbeam}
 \label{sec:testbeam}
 The first version of the CaRIBOu system has been deployed for the test of the \mbox{AMS180v4} sensors from November 2015 at the H8 beamline of the CERN Super Proton Synchrotron (SPS) which provides a 180 GeV$\slash$c $\pi^{+}$ beam. \par
 
 The CaRIBOu system is used to provide powers and bias voltages to the readout chip (\mbox{FE-I4B}) and the sensor undertest (\mbox{AMS180v4}), sensor configuration, sensor parameter tuning, power monitoring and logging.
 The \mbox{FE-I4B} ASICs are read out through the telescope readout system to facilitate the data analysis with the existing reconstruction software.
 The \mbox{FE-I4B} based telescope is also used for tracking and triggering \cite{fei4telescope}.\par
 
\begin{figure}[htbp]
\centering 
\includegraphics[width=0.7\textwidth]{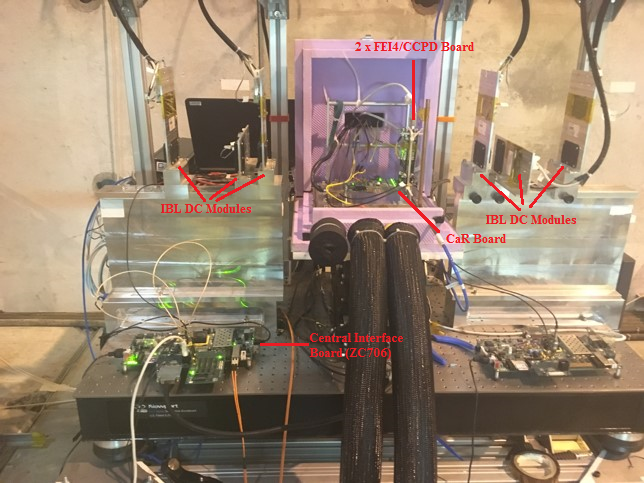}
\caption{\label{fig:beam_test}Two FEI4 boards with CCPD boards plugged in the CaR board which is placed on the telescope platform and connected to the Xilinx ZC706 board with a VHDCI adapter card and cable.}
\end{figure}

A photograph of the testbeam setup, with the boards housing the sensors and its readout \mbox{FE-I4B} ASIC plugging in the CaR board, which is mounted on the \mbox{FE-I4B} based telescope platform, is shown in Figure~\ref{fig:beam_test}. \par

The overall measured efficiency of the non-irradiated \mbox{AMS180v4} sensor exceeds 99.5\%. Details on the experimental setup and the analysis results can be found in \cite{2015testbeamresults}.
 
\section{Conclusion}
\label{conclusion}
In this paper, the design of the CaRIBOu modular system for characterizing silicon pixel detectors has been described. Five custom boards have been designed to test \mbox{AMS180v4} sensors glued to \mbox{FE-I4B} readout chips. Two data and command links have been implemented with Gigabit Ethernet link and GBT optical link. The CaRIBOu system has been employed successfully in the testbeams for the sensor AMS18v4. This modular test system has already been used for the testbeam for the HV-CMOS large-surface sensor demonstrator H35DEMO with an upgraded CaR board and front-end chip board in the 2016 testbeam at the CERN SPS. It is a versatile system that can be adapted to test other silicon detectors.
 
\acknowledgments
We thank Dr. S. Kulis at CERN, for providing the design files and documentation of the uASIC. We also thank Dr. M. Backhaus at CERN and J. Liu at CPPM, for the informative discussion during the hardware design stage of CaRIBOu. We would like to thank D. Pinelli at BNL for his excellent wire bonding service. We also acknowledge the help from A. Hoffmann and K. Sexton at BNL for the mechanical assembly of CaRIBOu.



\end{document}